\documentclass[12pt]{article}

\usepackage{times}

\usepackage{amsmath,amssymb}
\usepackage{graphicx}

\usepackage{xcolor}

\usepackage{hyperref}

\usepackage{soul}

\newcommand{\pin}{\psi_{\text{in}}}
\newcommand{\pout}{\rho_{\text{out}}}
\newcommand{\Hh}{\mathcal{H}}
\newcommand{\Tr}{\mathrm{Tr}\,}
\newcommand{\Hout}{\Hh_{\text{out}}}
\newcommand{\Hin}{\Hh_{\text{in}}}
\newcommand{\E}{\mathcal{E}}

\title{\vspace*{-0.5cm} Probing the limits of quantum theory\\ with quantum information at subnuclear scales}

\author{
Micha{\l} Eckstein$^{1,2 \star}$, Pawe{\l} Horodecki$^{3,4 \dagger}$
\\
\normalsize{$^{1}$ Institute of Theoretical Physics, Jagiellonian University,}\\
\normalsize{ul. {\L}ojasiewicza 11, 30--348 Krak\'ow, Poland}\\
\normalsize{$^{2}$ Copernicus Center for Interdisciplinary Studies,}\\
\normalsize{ul. Szczepa\'nska 1/5, 31-011 Krak\'ow, Poland}\\
\normalsize{$^{3}$ International  Centre  for  Theory  of  Quantum  Technologies,  University  of  Gda\'nsk,}\\
\normalsize{Wita  Stwosza  63,  80-308  Gda\'nsk,  Poland}\\
\normalsize{$^{4}$ Faculty of Applied Physics and Mathematics, National Quantum Information Centre,}\\
\normalsize{Gda\'nsk University of Technology, Gabriela Narutowicza 11/12, 80-233 Gda\'nsk, Poland}\\
\vspace*{0.8cm}
\normalsize{$^\star$e-mail:  michal.eckstein@uj.edu.pl, $^\dagger$e-mail: pawhorod@pg.edu.pl}
}

\date{\today}

\begin{document}

\maketitle 

\begin{abstract}
Modern quantum engineering techniques enabled successful foundational tests of quantum mechanics. Yet, the universal validity of quantum postulates is an open question. Here we propose a new theoretical framework of Q-data tests, which recognises the established validity of quantum theory, but allows for more general --- `post-quantum' --- scenarios in certain physical regimes. It can accommodate a large class of models with modified quantum wave dynamics, correlations beyond entanglement or general probabilistic postulates. We discuss its experimental implementation suited to probe the nature of strong nuclear interactions. In contrast to the present accelerator experiments, it shifts the focus from high-luminosity beam physics to individual particle coherent control.
\end{abstract}

\section{Introduction}

Quantum mechanics is one of the most successful scientific theories of the 20th century, faithfully modelling phenomena in the micro-world. The manifestation of some of its most distinctive features --- entanglement \cite{Horodecki}
 and wave-particle duality 
 \cite{C60}
--- require precise preparation of the system's state and detection of individual particles. Suitable devices for quantum engineering, based on electromagentic interactions, 
became available only recently. On the theoretical side, the possibility of acute control of quantum states gave birth to the theory of quantum information \cite{NielsenChuang}.
The recognition of entanglement and coherence as resources 
\cite{QResources}
leads to tantalising technological perspectives, including quantum computation, 
\cite{QuantumComp2016}
quantum cryptography \cite{QCryptography}
and quantum sensing \cite{QSensing}.
In parallel, quantum field theory emerged from the unification of quantum mechanics with the special theory of relativity \cite{Weinberg}. 
It is at the core of the Standard Model of particle physics and provides an extremely accurate framework for the study of high-energy phenomena.

The tremendous success of quantum theory motivates a question about its universality and limits of validity. Could there be a `post-quantum' theory violating some of the basic quantum principles? If so, in which physical regime would it become manifest? These questions have been approached from a number of different standpoints. One of them, outlined already in the 1960 by Louis de Broglie \cite{deBroglie1960}, assumes a nonlinear modification of the Schr\"odinger equation \cite{IBB_nonlinear,WeinbergNQM}, possibly along with a revision of the Born rule \cite{Czachor1998,Gisin2001}. A related class of theories seeks an objective mechanism behind the collapse of the quantum wave function \cite{OR}. A distinct strategy, developed more recently, is based on the possibility of non-local correlations stronger than those predicted by quantum mechanics \cite{PopescuRohrlich94,Bell_Nonlocal,AlmostQ,PawelRaviCausality}. Yet a different route leads through the axiomatisation of quantum theory in purely operational terms, which opened up a broader framework of so-called General Probabilistic Theories (see \cite{Chiribella16} and references therein).

It is usually expected that if there are any deviations from the standard quantum theory, then they might be related to the nature of the gravitational field \cite{OR}. This assumption points to two physical regimes of interest. The first one is determined by extremely short distances of the order of Planck length $1.6 \cdot 10^{-35}$~m or exceedingly large energies around the Planck energy $1.2 \cdot 10^{19}$ GeV \cite{DSR,QGPhenomenology,QG_optics}. The second regime involves quantum superposition of macroscopic objects of size $\gtrsim10^{-6}$ m and mass $\gtrsim10^6$ GeV/$c^2$ \cite{OR,LimitsQM}.
No experiment probing either of these domains has so far hinted at any new physics beyond the standard quantum theory \cite{Holometer_res2016,Superpos2019,Diosi21}.

\pagebreak

Yet, there exists another physical regime which may hide surprises. It is the interior of the nucleon, governed by the strong nuclear force, characterised by length scales of the order of $10^{-15}$~m and energies in the GeV range. Its key properties are well understood within the established theory of quantum chromodynamics (QCD). The latter draws a complex picture of the nucleon consisting of three valence quarks immersed in a sea of gluons and virtual quark--antiquark pairs \cite{Nucleon2001}. This makes the nucleon an intriguing quantum system, very dissimilar to the ones encountered in atomic physics. In particular, the natural expectation that the spins of the three constituent quarks add up to the nucleon's spin fails. The actual spin structure of the nucleon is much more convoluted, with the gluons' total angular momentum contributing up to 50\% of the total spin budget \cite{Bass2005,Bass2013,Spin21}.

These peculiar features of the nucleon motivate some fundamental questions: What are the intrinsic degrees of freedom at subnuclear scales? Are those degrees of freedom effectively classical, quantum or maybe post-quantum?  How is the information distributed and processed within nucleons?  Given the significant role played by the gluons and sea quarks, one should expect a high accumulation of degrees of freedom in a tiny volume. A typical high-energy experiment, involving $\sim 10^{11}$ particles, does not probe the individual subnucleonic degrees of freedom, but rather some collective features. One could thus speculate that the hypothetical post-quantum nature of individual quarks and gluons might be concealed by abundant pair creation processes dominating at high energies. Consequently, we claim that new insights into the nature of the strong interaction could be gained via precise quantum engineering experiments at the few-particle level.

However, this cannot be achieved within the standard prepare-and-measure paradigm at the level of individual particles because quarks and gluons
do not exist as free particles, but are always confined within hadrons. Therefore, one cannot implement the standard scenarios involving preparation and measurement of individual particles, akin to the Bell's test, which could detect super-quantum correlations \cite{Bell_Nonlocal}. For the same reason, it is impossible to carry out an interferometric experiment, which would probe the possible deviations from linear quantum dynamics \cite{OR}.

We show, however, that hypotheses about information processing at subnuclear scales can be rigorously formulated and answered in future precision experiments. To this end, we establish a new `black box' framework with quantum inputs and outputs. It is universal, model-independent, and can accommodate a large class of different post-quantum scenarios.
We demonstrate on concrete examples how the protocols can serve to detect phenomena, which could not be modelled within the standard quantum theory. Our approach is thus radically different from the existing studies (see e.g. \cite{Kaons2012,
BK_ent,EPR_quarks,Ent_weak,QI_top} and references therein) devoted to witness the genuinely quantum feature of entanglement in high-energy phenomena.

In the second part of the article 
we discuss some key elements needed for the implementation of our theoretical scheme in future experiments. The basic idea involves an orchestrated scattering of `quantum-programmed' projectiles from a target followed by projective measurements of a chosen observable on individual outgoing particles. In this way one effectively performs a quantum tomography of an unknown process  \cite{PTomography1,PTomography2}. More advanced experimental schemes aimed at probing the strength of correlations would involve multiple, subsequent or simultaneous, elastic scatterings from the same nucleon. These would require acute control of single nucleons and precise projective measurements of individual projectiles with energies in the GeV range, which poses serious technical challenges. We argue, however, that some of the foundational tests could be implemented in the state-of-the-art scattering experiments involving polarized beams, through meticulous quantum state reconstruction from the acquired data.

\section{\label{sec:theo}The theory-independent framework}

An elemental physical experiment
results in a set of conditional probabilities $\{P(a_j \, \vert \, x_i)\}_{i,j}$, calculated via the standard frequency method, for given input data $\{x_i\}_i$ and registered outcomes $\{a_j\}_j$. This fact is a starting point for the ``theory-independent'' paradigm  \cite{Bell_Nonlocal}, which focuses exclusively on the effective outcome probabilities, while ignoring the description of physical details of the studied systems. In such an approach the physical system under investigation is seen as a ``black box'', which can be probed with programmable input and controllable output systems. The latter carry the encoded information ($x_i$ at the input and $a_j$ at the output), while the box acts as an information processing device. Different theoretical schemes 
impose different constraints on the admissible information processing protocols executed by the box, hence lead to different predictions about the conditional probabilities $P(a_j \, \vert \, x_i)$, which can be directly registered in a suitable experimental scenario.

The archetypical example of such an experiment is the Bell test, in which two space-like separated parties --- Alice and Bob --- perform a measurement on a particle from an entangled pair, following a choice of the measurement setting. Hence, in a Bell test the measurable probabilities take the form $P(a,b \, \vert x, y)$, with  $a,b \in \{-1,+1\}$ and $x,y \in \{0,1\}$ denoting the outputs and inputs, respectively ($(a,x)$ for Alice and $(b,y)$ for Bob). It is well-known \cite{CHSH} that a wide class of ``local hidden variables'' theories implies a bound on the quantity $S = C(0,0) + C(0,1) + C(1,0) - C(1,1) \leq 2$ , where $C$ is the correlation function defined as $C(x,y) = \sum_{a,b} ab \, P(a,b \, \vert x, y)$. Multiple experiments have shown (see \cite{AspectBell2015} and references therein) that this inequality is violated in Nature and pointed to the validity of the quantum mechanical description, which predicts \cite{Cirelson} the bound $S \leq 2 \sqrt{2}$.

Here we put forward a conceptually new theory-independent framework founded on quantum information. We regard a physical system under investigation as a \emph{quantum-information} processing device, which could be probed with controllable quantum systems. The dynamics of quantum information effectuated by such a `quantum-data black box' (\emph{Q-data box}, for short) can be modelled according to different theoretical schemes. We set only two general constraints on the admissible theoretical description of a Q-data box:

\begin{enumerate}

	\item[(LO)]\label{LO} \textit{Locality}: A Q-data box is physically bounded in space.

	\item[(NS)] \textit{No-signalling}: A Q-data box cannot facilitate superluminal communication of any information.

\end{enumerate}

The demand of locality is a basic methodological assumption, which guarantees that the Q-data box is an operational notion. The boundedness in space means that any Q-data box can, in principle, be probed in isolation (i.e. ``shielded'') from the rest of the Universe. Let us emphasise that the locality of a Q-data box constrains only the admissible interaction with it and not the effectuated dynamics of quantum information (see Subsections \ref{sec:tomo} and \ref{sec:Helstrom}).  

The (NS) condition is a standard one and assures basic compatibility with relativity \cite{Bell_Nonlocal}. Here we take it in the broadest sense of \cite{PawelRaviCausality}, which allows for scenarios involving non-local dynamics of correlations.

\begin{figure}[h]
\begin{center}
\includegraphics[scale=1]{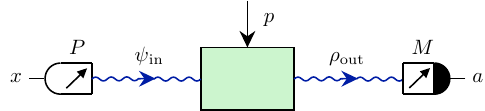}
\caption{\label{fig:Qbox}The basic scheme of a Q-data test. The green rectangle illustrates a Q-data box, black straight lines depict the classical information and the blue curved ones the quantum information. The vertical arrow over the top signifies that $p$ are classical parameters of the test.}
\end{center}
\end{figure}

In order to unveil the properties of a given Q-data box one performs a \emph{Q-data test}, which consists in probing a Q-data box with systems carrying the encoded quantum information (see Fig. \ref{fig:Qbox}). The relevant input and output data are now quantum states $\pin$, $\pout$ defined, respectively, over finite-dimensional Hilbert spaces $\Hh_{\text{in}}$ and $\Hh_{\text{out}}$. This constitutes a valid experiment when combined with an orchestrated initial quantum state preparation $P: x \to \pin$ and a final projective measurement of a chosen observable $M: \pout \to a$. Furthermore, the test can involve some classical parameters $p \in \Omega$, e.g. the energy of the ingoing system, the scattering angle or the direction of a tunable global magnetic field.


\pagebreak

The operationality of a Q-data test hinges upon three assumptions:

\begin{itemize}

	\item[(Prep)] \textit{Preparation}: The input state $\pin$ is pure and can be prepared arbitrarily precisely.
	
	\item[(Tomo)] \textit{Tomography}: The output state $\pout$ can be reconstructed to an arbitrary precision.
	
	\item[(Free)] \textit{Freedom of choice}: The initial state $\pin$ and the parameters $p$ can be chosen freely, i.e. independently of the state of the probed Q-data box.	

\end{itemize}

The first two conditions are guaranteed if we recognise the validity of quantum mechanics outside of the Q-data box, whereas the ``freedom of choice'' is a standard assumption adopted in the black-box approach \cite{PTomography1}.

It is vital to stress the operational difference between the input and output quantum data. Whereas assumptions (Prep) and (Free) guarantee that the input state $\pin$ is fully under control, the output $\pout$ is only an \emph{effective} quantum state reconstructed from the measurement outcomes $\{a_j\}_j$ via the quantum state tomography \cite{Tomo2004}. The latter is an algorithmic method for estimating an unknown quantum state from multiple repetitive measurements of observables $M_i$ from a tomographically complete set $\{M_i\}_{i=1}^{n^2}$  on $\Hh_{\text{out}}$, where $n = \dim \Hh_{\text{out}}$. Consequently, $\pout$ is in general expected to be mixed. Its impurity can be a combined effect of  imperfections of the experimental tomography and an objective indeterminacy caused, e.g. by the entanglement of the output system with the Q-data box. If $N$ denotes the number of measurements of each observable $M_i$ and $\pout^{[N]}$ is the corresponding reconstructed state, then assumption (Tomo) guarantees that $\pout^{[N]}$ converges in the limit $N \to \infty$ to a unique mixed state over the Hilbert space $\Hout$.

In summary, a Q-data test yields a dataset of the form $\{\pin^{(k)}, p^{(\ell)}; \pout^{(k,\ell)}\}_{k,\ell}$ with the indices $k$, $\ell$ ranging over the input states from $\Hh_{\text{in}}$ and parameters from the set $\Omega$, respectively. For every fixed input state $\pin^{(k)}$ and parameters' values $p^{(\ell)}$ one needs to complete the quantum state reconstruction, which yields an effective state $\pout^{(k,\ell)}$. The more tomographic measurements $N$ are performed for each input data $\pin^{(k)}, p^{(\ell)}$, the more credible the output quantum data is.

\begin{figure}[h]
\begin{center}
\includegraphics[scale=1]{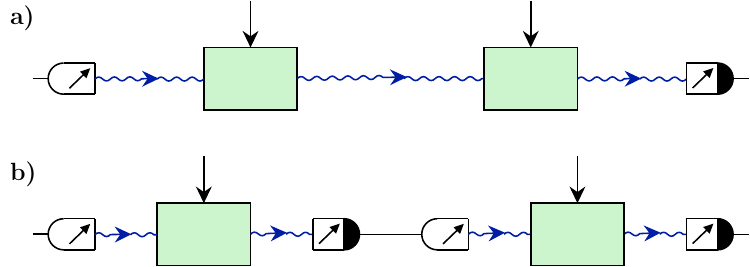}
\caption{\label{fig:concat}The scheme depicted in figure \textbf{a)} constitutes a valid Q-data test of a concatenation of two Q-boxes, but it is, in general, not equivalent to the concatenation of Q-data tests depicted in figure \textbf{b)}. This is because the system exiting the first Q-data box will typically be correlated with the first Q-data box. In scenario \textbf{a)} these correlations may affect the behaviour of the second Q-data box. In contrast, in scenario \textbf{b)} the second Q-data test can only depend upon the effective output state $\pout$ of the first test, which does not keep track of the correlations between the outgoing system and the first box.}
\end{center}
\end{figure}

An implementation of a Q-data test may involve a single system, prepared in an initial state $\pin$, which interacts with a given Q-data box, e.g. via scattering, and is then subject to the final tomographic measurement. In different scenarios, e.g. involving absorption and subsequent emission, the outgoing system will not be the same as the ingoing one. In either case, one should keep in mind that the outgoing system will typically be correlated with the probed Q-data box. In consequence, whereas the concatenation of Q-data boxes can be probed within a single Q-data test, the latter will \emph{not}, in general, be equivalent to the concatenation of Q-data tests probing individual Q-data boxes (see Fig. \ref{fig:concat}).

\begin{figure}[ht]
\begin{center}
\includegraphics[scale=1]{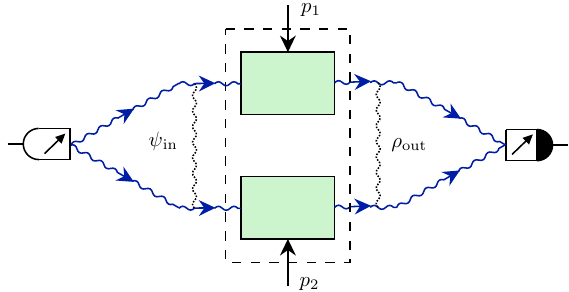}
\caption{\label{fig:Qbox2}An illustration of a non-local Q-data test involving two Q-data boxes in parallel. The dashed rectangle signifies that the local Q-data boxes can be correlated. The dotted curved lines point to the fact that both states $\pin$ and $\pout$ can be entangled.}
\vspace*{0.5cm}
\includegraphics[scale=1]{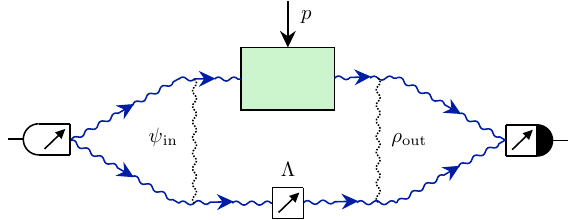}
\caption{\label{fig:Qbox3}A Q-data box can be probed with a quantum system entangled with a reference system. Because every local quantum device $\Lambda$ is naturally a Q-data box, this scenario is a particular case of a non-local Q-data test.
}
\end{center}
\end{figure}

Whereas Q-data boxes are --- by assumption (LO) --- local, the Q-data tests need not be so. The non-locality of quantum data allows one to probe multiple Q-data boxes `in parallel' using engineered input states on a product Hilbert space $\Hh_{\text{in}} = \otimes_n \Hh_{\text{in}}^n$ and a joined tomography over $\Hh_{\text{out}} = \otimes_{n} \Hh_{\text{out}}^n$  (see Figure \ref{fig:Qbox2}). The involved Q-boxes can be correlated, e.g. as a result of a preceding interaction, and a non-local Q-data test can detect these correlations. Non-local Q-data tests can also be performed with entangled input states.

A particular instance of a non-local Q-data test arises when a given Q-data box is probed locally by a quantum system entangled  with a reference system (see Fig.~\ref{fig:Qbox3}). In such a case, the total initial state has the form $\pin = \sum_i c_i \chi^i_B \otimes \eta^i_R$, where $B$ denotes the system input into the box, $R$ is the one kept outside and $c_k$ are complex numbers determining the correlations between $B$ and $R$. The final state will have an analogous form: $\pout = \sum_j d_j \sigma^j_B \otimes \xi^j_R$. In such an extended scenario, the full structure of $\pout$, including the mutual phases between local states, is reconstructed via quantum tomography with the help of projective measurements from the tomographically complete basis on $\Hh_\text{out} = \Hh^B_\text{out} \otimes \Hh^R_\text{out}$.

Finally, let us explain why in a Q-data test one is allowed to use only pure states at the input. The reason is essentially the same as in the context of a standard experiment involving classical data: In a single experimental run the input is always definite --- its indeterminacy would basically mean that we are unsure of what the input was, hence the experiment was flawed by the introduction of subjective uncertainty. Similarly, the preparation of a mixed quantum state effectively amounts to introducing uncontrolled entanglement with an environment \cite{Zurek_decoh}. Consequently, probing a Q-data box with a quantum system prepared in a mixed state is in fact a non-local Q-data test of the form described in Fig. \ref{fig:Qbox3} with the reference system being out of control.

Let us now turn to concrete examples of Q-data tests designed to assess the validity of different post-quantum scenarios.

\section{Examples of Q-data tests}

\subsection{\label{sec:tomo}Quantum process tomography}

From the perspective of Q-data tests, a Q-data box effectuates some quantum-information processing protocol described, for any fixed values of external parameters' $p \in \Omega$, by a map
\begin{align*}
\E: \Hin \to S(\Hout),
\end{align*}
where $S(\Hout)$ denotes the space of density operators on $\Hout$. Given a dataset collected during a Q-data test $\{\pin^{(k)}, p; \pout^{(k)}\}_{k}$ one can attempt to reconstruct this map.

Quantum mechanics imposes rather tight constraints on the admissible form of $\E$. Concretely, if the Q-data box operates according to the quantum principles, then $\E$ must be a linear completely positive trace preserving (CPTP) map \cite{NielsenChuang}. Such a map is completely determined by $m^2 (n^2-1)$ real parameters with $m = \dim \Hin$, $n = \dim \Hout$ and extends uniquely to a map $\widetilde{\E}: S(\Hin) \to S(\Hout)$. These parameters can be directly measured in the quantum process tomography scheme \cite{PTomography1}. The latter consists in probing the box with $m^2$ different pure input quantum states $\pin^{(k)} \in \Hin$, which form a basis of the space $S(\Hin)$, and reconstructing the corresponding output states $\pout^{(k)}$.

Let us fix such a basis $\{\pin^{(k)}\}_{k=1}^{n^2}$ and consider its unitary rotation $\{U_\delta (\pin^{(k)})\}_{k=1}^{n^2}$ with some tunable parameter $\delta$. Let $\E_\delta$ denote the corresponding quantum channel, which is reconstructed from the gathered data $\{U_\delta(\pin^{(k)}), p; \pout^{(k)}\}_{k}$. If the probed Q-data box abides by the laws of quantum mechanics then the outcome of the process tomography does not depend on the choice of the basis for input states, that is $\E_\delta = \E_{\delta'}$ for all $\delta, \delta'$.

A dependence of the reconstructed map $\E_\delta$ on the parameter $\delta$ would provide evidence for the post-quantum nature of information processing within the probed Q-data box. Such a deviation could be quantified with the help of any standard distinguishability measure \cite{Zyczkowski}, e.g. the quantum fidelity between the Choi--Jamio{\l}kowski matrices of $\E_\delta$ and $\E_{\delta'}$.

A different scenario of a Q-data test exploits an alternative, ancilla-assisted, quantum process tomography scheme \cite{AATomo}. In the latter, the Q-data box is probed with a quantum system entangled with a reference system, as in Fig. \ref{fig:Qbox3}. Now, there is a single fixed input state $\pin$ and the map $\E$ is reconstructed from joint measurements on the system exiting the Q-data box and the reference system. If $\E$ is a CPTP map then the two schemes of quantum process tomography are equivalent. Consequently, one can probe a Q-data box by experimentally checking this equivalence.

The General Probabilistic Theories beyond quantum mechanics allow for the dynamics of states, which does not fulfil the CPTP condition \cite{MarkovGPT}. Such a tomographic Q-data test could thus serve to test quantum theory from a general probabilistic perspective.

\subsection{\label{sec:Helstrom}The Helstrom discrimination test}

Consider an Alice producing one of the two quantum states 
$\psi_{1}, \psi_2$ with the corresponding probabilities $p_1, p_2$.  Bob's task is to discriminate between the two inputs in an optimal way. The probability of success is defined as
\begin{equation}
P_{\text{succ}}( p_1, \psi_1, p_2, \psi_2) := \sum_{i=1}^{2} p_i \, P(a=i|\psi_i),
\label{Probability-correct}
\end{equation}
where the result ``$a=1$'' (``$a=2$'') corresponds to the 
a posteriori conclusion drawn by Bob ``Alice prepared the system 
in a state $\psi_1$ ($\psi_2$)''. 

Suppose that Bob is capable of executing any unitary quantum dynamics on the given state and performing any projective measurement on the final state. His probability of success in discriminating between the two quantum states is limited by the Helstrom bound \cite{Helstrom_book}:
\begin{equation}
P_{\text{succ}}( p_1, \psi_1, p_2, \psi_2)
\leq \tfrac{1}{2} \big(1 + \Tr \vert p_1 \psi_1 - p_2 \psi_2 \vert \big).
\label{Helstrom}
\end{equation}
The latter is strictly smaller than 1, unless the two states $\psi_1, \psi_2$ are orthogonal.

Suppose now that Bob probes a Q-data box with the obtained state $\psi_i$ and registers a state $\rho'_i$ at the output. If, using $\rho'_i$, he can exceed the Helstrom bound \eqref{Helstrom}, then the Q-data box must have effectuated some post-quantum process. Observe that even if the outgoing states $\rho'_i$ have lower purity than the ingoing $\psi_i$, they might still provide better distinguishability.

The violation of the Helstrom bound \eqref{Helstrom} is a generic feature of nonlinear modifications of the Schr\"odinger equation \cite{Childs}. However, it is well known \cite{Gisin1989} that nonlinear quantum dynamics leads to superluminal signalling and hence violates one of our basic assumptions (NS) about admissible models of Q-data boxes. This happens when Alice sends to Bob a quantum system, which is maximally entangled with another one kept at her local laboratory, say $\Psi_{AB}^- = \tfrac{1}{\sqrt{2}} \big( \vert 0 \rangle \vert 1 \rangle - \vert 1 \rangle \vert 0 \rangle \big) = \tfrac{1}{\sqrt{2}} \big( \vert + \rangle \vert - \rangle - \vert - \rangle \vert + \rangle \big)$. By effectuating a projective measurement in one of the bases $\{0,1\}$ or $\{+,-\}$ on her system Alice can prepare one of the two statistical ensembles $\{ \tfrac{1}{2} \vert 0 \rangle ; \, \tfrac{1}{2} \vert 1 \rangle \}$ or $\{ \tfrac{1}{2} \vert + \rangle ; \, \tfrac{1}{2} \vert - \rangle \}$ entering Bob's laboratory. These ensembles yield \emph{the same} quantum density matrix, but nonlinear quantum dynamics generically maps them into another pair of ensembles, which result in two \emph{different} density matrices \cite{Gisin2001,Gisin1989}. Consequently, Bob could distinguish these two situations hence immediately learning Alice's decision at a distance.

One way to save the model's consistency is to modify the static structure of quantum mechanics \cite{Gisin2001}. In our approach this is not allowed, because we assume the validity of quantum mechanics outside of the Q-data box. Therefore, one would have to assume that the Q-data box introduces some stochastic element \cite{OR}, for instance by effectuating a spontaneous von Neumann collapse, which destroys the entanglement without modifying Alice's local density matrix.

An experimental violation of the Helstrom bound with correctly prepared pure input states $\psi_1, \psi_2$ would provide strong evidence for the breakdown of quantum theory in the probed Q-data box. In such a case, further non-local Q-data tests involving a controlled reference system might help unveiling the correct post-quantum model.

\subsection{\label{sec:NQRAC}Quantum random access code boxes}

\begin{figure}[h]
\begin{center}
\includegraphics[scale=1]{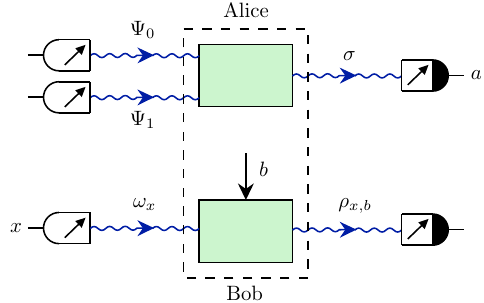}
\caption{\label{fig:NQRAC}An example of a quantum random access code box involving two correlated Q-data boxes with post-selection of events $a=b$ (see text for the description).}
\end{center}
\vspace*{-0.45cm}
\end{figure}

Let us now present an example of a Q-data test designed to probe the strength of correlations. It exploits a quantum random access code box involving both classical and quantum data \cite{Grudka2015}. The relevant Q-data box has three quantum inputs ($\Psi_0$ and $\Psi_1$ for Alice and $\omega$ for Bob) and two quantum outputs ($\sigma$ for Alice and $\rho$ for Bob), along with a two-bit external parameter $b$ on Bob's side (see Fig. \ref{fig:NQRAC}). The final measurement on Alice's side yields four classical outcomes enumerated by a pair of bits $a$. For sake of concreteness, let us assume that the involved quantum states are qubits.

In the task of quantum random access code Bob wants to learn one of the Alice's qubits with his choice being parametrised by a random bit $x$. Alice sends to Bob her 2-bit classical output $a$. Using these as his input $b$, Bob steers his initial state $\omega_x$ into $\rho_{x,b}$. The task is successful if Bob's quantum output $\rho_{x,b=a}$ is equal to $\Psi_{x}$ for both choices $x=0$ and $x=1$. In an equivalent protocol depicted in Fig. \ref{fig:NQRAC} one generates 2 random bits $b$ and post-selects for events with $a=b$.

In \cite{Grudka2015} it was shown, that such a task can never be achieved if the Q-data box works according to the rules of quantum mechanics. On the other hand, if it involves some post-quantum correlations --- the Popescu--Rohlich no-signaling boxes \cite{PopescuRohrlich94} --- Bob can always recover perfectly the qubit of his choice.

The success rate of this quantum random access code can be quantified using the quantum fidelity $F$ (ref. \cite{Fidelity})
\begin{align}
P_\text{succ} = \tfrac{1}{2}\sum_{x=0,1} F( \Psi_x, \rho_{x,b=a} ).
\end{align}

Quantum mechanics imposes \cite{EHS2020} an explicit bound  $P_\text{succ} \leq 3/4$, valid for any choice of the input states $\Psi_x$. If in a experiment we achieve a success rate exceeding $3/4$, then we know that the Q-data box must involve some post-quantum element.

Let us note that the presented scheme for testing the strength of correlations is conceptually different than the classical Bell--CHSH test \cite{CHSH}, which also admits post-quantum scenarios \cite{Bell_Nonlocal}. Our test \emph{assumes} the validity of quantum mechanics at the input and at the output, but admits post-quantum imprints on the input-output correlations between $\pin$ and $\pout$. Similar tests could involve, for instance, post-quantum steering scenarios, in which one or more parties collaborate to generate a desired quantum state at receiver's output  \cite{Postquantum_steering,Postquantum_steering2}.

Typically, in experiments designed to test the strength of correlations, such as the Bell test, one assumes that the two parties, Alice and Bob, are spacelike separated \cite{Bell_Nonlocal}. Indeed, if Alice and Bob are close enough to interact and exchange information during the experiment, then any correlations between them can be simulated within the conventional classical + quantum paradigm. In the context of physics at subnuclear scales the spacelike separation of the inputs is not achievable. Nevertheless, any deviation from quantum bounds on correlations, such as the one presented above, would signify a highly peculiar dynamics of information, not realised naturally in any known composite quantum system.

\subsection{\label{sec:QNS}Quantum no-signalling boxes}

\begin{figure}[h]
\begin{center}
\includegraphics[scale=1]{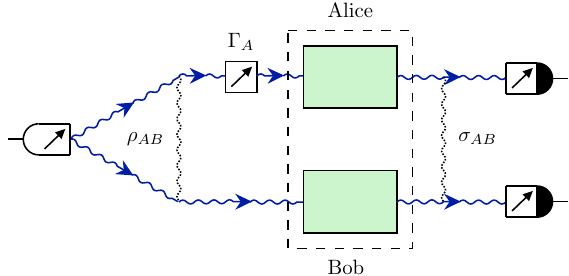}
\caption{\label{fig:QNS}A Q-data test incarnating a quantum no-communication scenario (see text for the description). The dashed rectangle signifies that Alice's and Bob's Q-data boxes are correlated. The dotted curved lines point to the fact that both states $\rho_{AB}$ and $\sigma_{AB}$ can be entangled.}
\end{center}
\vspace*{-0.4cm}
\end{figure}

In the previous examples the input quantum states were uncorrelated. One can also conceive more general scenarios, in which both the input and the output are entangled states. In this context one can ask what are the characteristic features of the dynamics of correlations imposed by the quantum theory. One specific example, the so-called \emph{no-signalling quantum boxes} (NSQ) -- see Fig.~\ref{fig:QNS}, has been analysed in \cite{QNS}. In this scenario one considers a special class of quantum maps on bipartite quantum inputs. A map $\Lambda_{AB}: \rho_{AB} \to \sigma_{AB}$ satisfies the condition of no-communication from Alice to Bob if for a given state $\rho_{AB}$ it maps the family of the inputs
$[\Gamma_{A} \otimes \text{id}_{B}](\rho_{AB})$, with any local unitary map $\Gamma_{A}$, into outputs $\sigma_{AB}(\Gamma_{A},\rho_{AB})$  
having {\it the same} Bob's reduced state $\sigma_{B} = \Tr_{\!A} \rho_{AB}$. An analogous condition of no-communication from Bob to Alice is assumed.

In \cite{QNS} it was proven that the set of such NSQ maps is of volume zero in the set of all bipartite dynamics. Since the trivial, i.e. non-interacting, dynamics is also of volume zero, a randomly chosen interacting quantum dynamics violates the no-communication condition. A relevant test would thus consist in the quantum process tomography with different randomly chosen initial data $\rho_{AB}$ and $\Gamma_{A}$. Actually, if one would be able to scan over all bipartite inputs $\rho_{AB}$, then $\Gamma_{A}$ would be redundant.

In the context of subnuclear physics one expects non-trivial interactions (and thus non-trivial dynamics of information), hence if the experiment reveals a significant percent of outcomes compatible with the no-communication condition, then it should be considered as an indication of a deviation from the standard quantum dynamics. Quantitatively, the typicality of communication in quantum dynamics can be estimated by means of uniform measures on CPTP maps~\cite{Random_channels}.

\section{Towards an experiment}

The theoretical scheme of Q-data boxes presented above does not depend on the physical context. It can equally well serve to test post-quantum scenarios involving, e.g. the gravitational field (cf. \cite{Marletto20,Rijavec20}).

We now sketch a general experimental setup for the implementation of a Q-data test in the context of strong nuclear interactions. To this end we regard free nucleons (or, more generally, hadrons) and nuclei as Q-data boxes, which can be probed with quantum information. These systems naturally meet the assumption (LO) of locality, because the strong nuclear force operates effectively only inside the nucleons and nuclei. Consequently, we assume that the hypothetical post-quantum phenomena take place inside nucleons and are negligible outside --- in accordance with the general framework of Q-data boxes. Let us recall that the assumption (LO) pertains only to the locality of interaction with a Q-data box and does not exclude correlations between spacelike-separated nucleons/nuclei.

We note that several successful implementations of quantum information processing protocols were carried out using \emph{collective} quantum degrees of freedom of nuclear ensembles \cite{NMR20,CollectiveNucl21}. These include i.a. a recent quantum interface between a single electron and a nuclear ensemble \cite{Interface_EN}. However, to probe the nature of strong nuclear interactions through Q-data tests one needs to address the \emph{internal} quantum degrees of freedom of nuclei. A successful implementation of a Q-data test involving the internal structure of nucleons would provide a direct test of the \mbox{(post-)quantum} nature of the strong nuclear interaction mediated by gluons. One can also envisage an analogous test for probing the residual strong interaction binding nucleons within a nucleus.

We propose that such implementations could be achieved through scattering of quantum-programmed projectiles on nucleonic or nuclear targets. Such a scenario guarantees the existence of natural classical parameters related to the kinematics of the collision.

\begin{figure}[h]
\begin{center}
\includegraphics[scale=1]{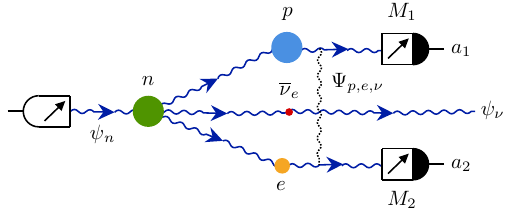}
\caption{\label{fig:nu}The conceptual scheme of preparation of a single neutrino quantum state. A free neutron $n$ prepared in a quantum state $\psi_n$ decays into a proton $p$, an electron $e$ and electron antineutrino $\overline{\nu}_e$. The quantum information encoded in $\psi_n$ is transferred into an entangled state $\Psi_{p,e,\nu}$ of the decay products. Through projective measurements on the proton and the electron one can steer the final neutrino state $\psi_\nu$, which depends on measurement settings $M_1, M_2$ and outcomes $a_1, a_2$.}
\end{center}
\vspace*{-0.5cm}
\end{figure}

In the first stage the state $\pin$ is prepared and imprinted on a projectile using standard quantum engineering techniques based on electromagetic interactions \cite{ph_prep,ph_prep_rev,e_prep,SQUIRRELS}. 
The carrier of the initial quantum information should preferably be a free electron or a photon, because of their controllability and stability, though one could in principle employ any fundamental particle, e.g. a positron or a muon. One might also conceive the preparation of the quantum state of an electron antineutrino through an orchestrated $\beta$-decay -- see Fig. \ref{fig:nu}.
A natural choice for the quantum degree of freedom encoding a qubit state $\pin$ is the electron's spin or photon's polarization. Alternative procedures might include the programming of particle's momentum, angular momentum or position quantum variables (wave packet shaping) \cite{PhaseShaping,PacketShaping}. 

During the second stage the prepared quantum state $\pin$ is input into the Q-data box via a precise scattering process. When the quantum-programmed electron or photon hits a nucleon it gets correlated, via the electromagnetic interaction, with some quantum system in the target --- a single quark or some conglomerate of them. The energy of the projectile should be large enough ($\gtrsim~1$~GeV), so that it probes the internal structure of the target and not some collective degree of freedom.

In the desirable case of elastic scattering there is a single outgoing projectile which carries the quantum information to be registered. In the more common inelastic collisions, the target disintegrates and the sought-for output quantum information is encoded in the global state of all the scattering products.

The quest for the final stage of the experiment is to perform a projective measurement of a chosen quantum degree of freedom on the outgoing projectiles. Suitable experimental schemes for projective measurements of the polarization states of individual $\gamma$-photons have recently been developed \cite{Gamma18,Gamma19}.
Furthermore, the measurement of spins of massive projectiles could be based on the quantum Stern--Gerlach scheme \cite{Espin99,QSternGerlach,Espin20}.

The type of the outgoing particle, together with the kinematic characteristics of the collision, and possibly other global features related e.g. to the polarization of the target, constitute the set $\Omega$ of classical parameters of the Q-data test. For every fixed value of these parameters, $p \in \Omega$, one needs to carry out a complete quantum tomography of the output state $\pout^p$ on the outgoing projectile. As explained in Section \ref{sec:theo}, this amounts to gathering the data of multiple measurements of different observables.
 If one expects some of the projectiles to be resulting from the same interaction vertex, then it is desirable to perform the quantum tomography of the joint state of these projectiles.

An idealistic Q-data test, akin to Bell-type experiments, should be performed with \textit{individual} particles. This is, however, a formidable task.

The first stumbling block is the preparation of a target consisting of a single nucleon or a single nucleus. Recently, a single proton has been isolated in a Penning trap for sake of measuring its magnetic moment \cite{Trap2017}. But a trap suitable for an implementation of a Q-data box should prepare a single nucleon in a sharp static position, to maximise the cross-section for the desired scattering process. This condition favours optical tweezers based on laser pulses \cite{Tweezer14},
 which, however, have not so far been engineered to trap single nucleons or nuclei.
 
The second major difficulty for such an idealistic Q-data test stems from the fact that a typical high energy collision   results in an abundance of the outgoing projectiles. Hence, one should expect that the sought for quantum information will be concealed in the entire collection of the decay products. In the Bell-test language this could be seen as a ``detection loophole'' (aka ``fair-sampling loophole'') \cite{Bell_Nonlocal}. In order to fathom out the input-output correlations probed in the Q-data test one should aim at clean experiments involving a manageable number of outgoing projectiles. Furthermore, the latter should be measured exclusively -- i.e. the detection should involve individual projective measurements on all projectiles.

\begin{figure}[hhh]
\begin{center}
\includegraphics[scale=1]{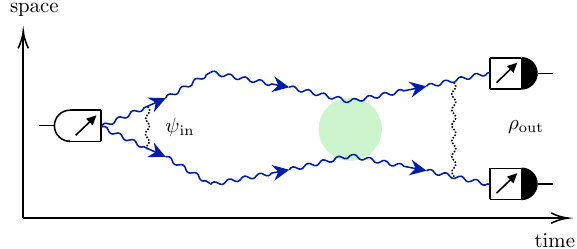}
\caption{\label{fig:space}The conceptual scheme for an implementation of a Q-data test aimed at probing the strength of correlations -- cf. Figs. \ref{fig:NQRAC} and \ref{fig:QNS}. It involves simultaneous scattering of two or more programmed quantum systems from the same Q-data box, followed by a joint projective measurement on the outgoing projectiles.}
\end{center}
\end{figure}

\begin{figure}[h]
\begin{center}
\resizebox{\textwidth}{!}{\includegraphics[scale=1]{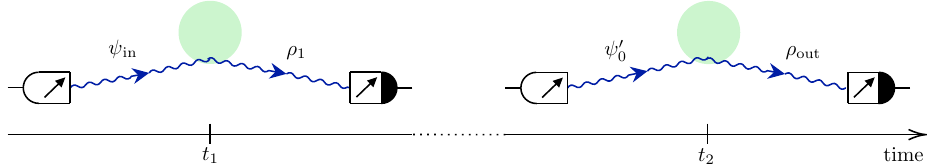}}
\caption{\label{fig:time}The conceptual scheme for an implementation of a Q-data test probing the dynamics of quantum information. At an initial time $t_1$ the input state $\pin$ is prepared and imprinted into the Q-data box. The state $\rho_1$ encoded in the scattered projectile is subject to a projective measurement shortly afterwards. At a later time $t_2$ the quantum information is read out from the Q-data box via a second scattering of the projectile carrying an `empty sheet' state $\psi_0'$. Finally, the quantum state $\pout$ is reconstructed from the tomographic measurements. Note that the first scattering process must be elastic, so that \emph{the same} Q-data box is probed at the second time-moment.}

\vspace*{0.8cm}

\includegraphics[scale=1]{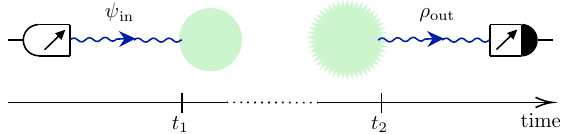}
\caption{\label{fig:AbsEm}An alternative scheme for probing the dynamics of quantum information. The projectile carrying the prepared state $\pin$ is absorbed into the Q-data box. Then, the box de-excites and decays into a few products. The final state $\pout$ is reconstructed from projective measurements on the outgoing projectiles.}
\end{center}
\vspace*{-0.4cm}
\end{figure}

The implementation of some of the Q-data test aimed at probing the strength of correlations, presented in Sections \ref{sec:NQRAC} and \ref{sec:QNS}, would require a coordinated multiple scattering from \emph{the same} nucleon/nucleus -- see Fig. \ref{fig:space}. In the same vein, one can envisage experiments probing the nature of time-like correlations, i.e. the dynamics of quantum information at subnuclear scales. Such a scenario requires at least two subsequent scatterings from \emph{the same} nucleon/nucleus (see Fig.~\ref{fig:time}), in contrast to the single-scattering scheme, which probes instantaneous interactions. An alternative scheme based on absorption and subsequent emission might be useful to probe \mbox{(post-)quantum} dynamical effects, the characteristic scales of which are very short -- see Fig.~\ref{fig:AbsEm}.


Such experiments are far beyond the available technologies. On the other hand, the Q-data tests presented in Sections \ref{sec:tomo} and \ref{sec:Helstrom} are conceivable with the state-of-the-art accelerator technology. Whereas quantum state engineering of individual high-energy particles poses a serious challenge, the polarized beams of electrons are routinely employed in accelerator experiments  \cite{Beams2005,E_beams21}. Some advances towards a generation of polarized positron beams was also made \cite{pos_pol1}. In parallel, a novel technique to generate highly-polarized multi-GeV photon beams has recently been proposed \cite{Gamma_pol1,Gamma_pol2}. Such a highly energetic polarized beam should then be scattered against a nucleonic or nuclear target. We note that protonic targets can also be polarized. The direction of the target's polarization can serve as an additional classical parameter of the test.

\begin{figure}[h]
\begin{center}
\includegraphics[scale=1]{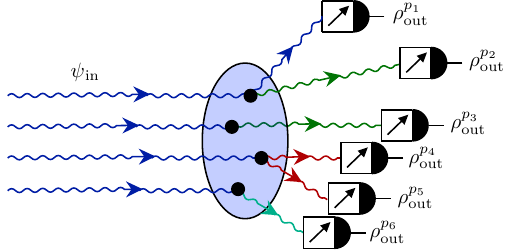}
\caption{\label{fig:beam}The scheme for an implementation of a Q-data test with polarized beams scattered on a nucleonic/nuclear target. The polarization of the ingoing beam fixes the initial quantum state $\pin$. At the output one performs projective measurements of a chosen degree of freedom, e.g. spin/polarization, of all of the outgoing projectiles. The states $\pout^p$ are reconstructed from the output data after post-selection with respect to the registered classical parameters pertaining to the kinematics of the collision and the species of the outgoing particles. 
The most promising region of the parameter space is that related to elastic scattering, because the output state is encoded in a single type of outgoing projectiles.
}
\end{center}
\end{figure}

The key challenge for such a beam-based Q-data test, as contrasted with typical accelerator experiments, is the accomplishment of precise quantum state tomography on the outgoing projectiles. This requires multiple experiments, with the same polarization of the ingoing beam, but different projective measurements of the spin/polarization (or other quantum degree of freedom) of the decay products. As explained earlier, one should post-select the gathered data to identify the scattering processes with different values of classical parameters, $p \in \Omega$, (particle species, scattering angle, energy etc.) and perform the reconstruction of $\pout^p$ for every value of~$p$ -- see Fig. \ref{fig:beam}.


The best chance to observe some non-trivial input-output correlations is associated with the events involving elastic scattering. This is because the entire available quantum information is carried by a single projectile, which one needs to measure projectively. The elastic scattering of electrons on protons with GeV energies has been performed at the Mainz Microtron MAMI \cite{MAMI10} and the Jefferson Lab \cite{ep_JefLab04,ep_JefLab19,ep_JefLab20}, with a good control of the inelastic background \cite{ep_JefLab19,ep_control10}. The quest would then be to carry out the quantum tomography on the outgoing electrons.

Within the adopted model-independent framework there is no hint about which processes could actually exhibit some post-quantum behaviour. We only postulate that the ingoing projectiles should have energy in the GeV range in order to address the internal degrees of freedom of the target. Also it is clear that the fewer types of the outgoing projectiles, the better the chance to observe non-trivial quantum information processing at subnuclear scales.
Otherwise, one should scrupulously explore the entire available space of classical parameters $p \in \Omega$. This blind search could be enhanced by the recently developed machine learning techniques for quantum experiments \cite{ZeilingerAI,ZeilingerAI2}.

\section{Discussion}

We have put forward a research programme designed to explore the information-theoretic properties of subnuclear phenomena. The adopted theoretical scheme allows one to probe the limits of quantum mechanics from an outside --- ``post-quantum'' --- perspective. Its implementation requires a new type of experiments, which combine the energy scales available in accelerator experiments with the precise quantum engineering of single particles. We have argued that some of the relevant Q-data tests could be designed basing on elastic scattering of polarized beams of electrons on a protonic target. The major technical challenge for the implementation of the proposed experimental scheme is associated with the need for performing tunable projective measurements of the spin of highly-energetic outgoing electrons.

Any evidence of a deviation from quantum-mechanical behaviour would have a profound impact on our understanding of fundamental physics, calling for a revision of the basic principles underlying the Standard Model of particles. Also, it is known that certain post-quantum theories can have dramatic consequences for information processing, e.g. yielding solutions to NP-complete problems in polynomial time \cite{NQM_NP} or trivialising communication complexity \cite{NTCC}. Consequently, the envisaged experimental programme provides an unprecedented opportunity for direct empirical tests of various physical principles \cite{NTCC,Aaronson2005,NANLC,InformationCausality}, postulated for sake of taming the undesired information-processing properties of Nature.

The exploration of quantum information protocols at subnuclear scales is also of significant interest from the conventional quantum perspective. Firstly, it provides an unparalleled test of the adequacy of the quantum-field-theoretic description of fundamental interactions, independent from the Lorentz and $CPT$ symmetries violation experiments \cite{LorentzCPT_2011}. Secondly, it would offer new possibilities for implementation of quantum information processing protocols at unprecedented scales and establish a solid ground for the quantum simulation of the nucleon's interior \cite{Lake2010,Kormos2017}.

\section*{Acknowledgements}

M.E. acknowledges the support of the Foundation for Polish Science under the project Team-Net NTQC no. 17C1/18-00. P.H. acknowledges support by the Foundation for Polish Science (IRAP project, ICTQT, contract no. 2018/MAB/5), co-financed by EU within Smart Growth Operational Programme.

We are particularly gratefully to Steven Bass for numerous inspiring discussions and pertinent comments on preliminary versions of the manuscript. We also thank Ryszard Horodecki and Tomasz Miller for valuable comments on the draft. We express our gratitude to the anonymous Referees, who helped us improve the manuscript. M.E. would also like to thank the Institute of Theoretical Physics and Astrophysics at University of Gda\'nsk and the National Quantum Information Center in Gda\'nsk for the hospitality.





\begin{thebibliography}{99}

\bibitem{Horodecki}
Horodecki R, Horodecki P, Horodecki M, Horodecki K. 2009  Quantum entanglement.
  {\em Reviews of Modern Physics} \textbf{81}, 865--942.

\bibitem{C60}
Arndt M, Nairz O, Vos-Andreae J, Keller C, Van~der Zouw G, Zeilinger A. 1999
  Wave--particle duality of $\mathrm{C}_{60}$ molecules. {\em Nature}
  \textbf{401}, 680--682.

\bibitem{NielsenChuang}
Nielsen MA, Chuang IL. 2010 {\em Quantum Computation and Quantum Information}.
Cambridge University Press.

\bibitem{QResources}
Chitambar E, Gour G. 2019  Quantum resource theories. {\em Reviews of Modern
  Physics} \textbf{91}, 025001.

\bibitem{QuantumComp2016}
Bartlett SD. 2016  Atomic physics: A milestone in quantum computing. {\em
  Nature} \textbf{536}, 35--36.

\bibitem{QCryptography}
Gisin N, Ribordy G, Tittel W, Zbinden H. 2002  Quantum cryptography. {\em
  Reviews of Modern Physics} \textbf{74}, 145--195.

\bibitem{QSensing}
Degen CL, Reinhard F, Cappellaro P. 2017  Quantum sensing. {\em Reviews of
  Modern Physics} \textbf{89}, 035002.

\bibitem{Weinberg}
Weinberg S. 1995 {\em The Quantum Theory of Fields}.
Cambridge University Press.

\bibitem{deBroglie1960}
de~Broglie L. 1960 {\em Non-linear Wave Mechanics: A Causal Interpretation}.
Elsevier.
Translated [from French] by Arthur J. Knodel and Jack C. Miller.

\bibitem{IBB_nonlinear}
Bia{\l}ynicki-Birula I, Mycielski J. 1976  Nonlinear wave mechanics. {\em
  Annals of Physics} \textbf{100}, 62--93.

\bibitem{WeinbergNQM}
Weinberg S. 1989  Testing quantum mechanics. {\em Annals of Physics}
  \textbf{194}, 336--386.

\bibitem{Czachor1998}
Czachor M. 1998  Nonlocal-looking equations can make nonlinear quantum dynamics
  local. {\em Physical Review A} \textbf{57}, 4122--4129.

\bibitem{Gisin2001}
Simon C, Bu\ifmmode~\check{z}\else \v{z}\fi{}ek V, Gisin N. 2001  No-Signaling
  Condition and Quantum Dynamics. {\em Physical Review Letters} \textbf{87},
  170405.

\bibitem{OR}
Bassi A, Lochan K, Satin S, Singh TP, Ulbricht H. 2013  Models of wave-function
  collapse, underlying theories, and experimental tests. {\em Reviews of Modern
  Physics} \textbf{85}, 471--527.

\bibitem{PopescuRohrlich94}
Popescu S, Rohrlich D. 1994  Quantum nonlocality as an axiom. {\em Foundations
  of Physics} \textbf{24}, 379--385.

\bibitem{Bell_Nonlocal}
Brunner N, Cavalcanti D, Pironio S, Scarani V, Wehner S. 2014  Bell
  nonlocality. {\em Reviews of Modern Physics} \textbf{86}, 419--478.

\bibitem{AlmostQ}
Navascu{\'e}s M, Guryanova Y, Hoban MJ, Ac{\'\i}n A. 2015  Almost quantum
  correlations. {\em Nature Communications} \textbf{6}, 1--7.

\bibitem{PawelRaviCausality}
Horodecki P, Ramanathan R. 2019  The relativistic causality versus no-signaling
  paradigm for multi-party correlations. {\em Nature Communications}
  \textbf{10}, 1701.

\bibitem{Chiribella16}
Chiribella G, Spekkens RW, editors. 2016 {\em Quantum Theory: Informational
  Foundations and Foils}.
Springer.

\bibitem{DSR}
Amelino-Camelia G, Ellis J, Mavromatos N, Nanopoulos D, Sarkar S. 1998  Tests
  of quantum gravity from observations of $\gamma$-ray bursts. {\em Nature}
  \textbf{393}, 763--765.

\bibitem{QGPhenomenology}
Amelino-Camelia G. 2013  Quantum-Spacetime Phenomenology. {\em Living Reviews
  in Relativity} \textbf{16}.

\bibitem{QG_optics}
Pikovski I, Vanner MR, Aspelmeyer M, Kim M, Brukner C. 2012  Probing
  {P}lanck-scale physics with quantum optics. {\em Nature Physics} \textbf{8},
  393--397.

\bibitem{LimitsQM}
Arndt M, Hornberger K. 2014  Testing the limits of quantum mechanical
  superpositions. {\em Nature Physics} \textbf{10}, 271--277.

\bibitem{Holometer_res2016}
Chou AS, Gustafson R, Hogan C, Kamai B, Kwon O, Lanza R, McCuller L, Meyer SS,
  Richardson J, Stoughton C, Tomlin R, Waldman S, Weiss R. 2016  First
  Measurements of High Frequency Cross-Spectra from a Pair of Large {M}ichelson
  Interferometers. {\em Physical Review Letters} \textbf{117}, 111102.

\bibitem{Superpos2019}
Fein YY, Geyer P, Zwick P, Kia{\l}ka F, Pedalino S, Mayor M, Gerlich S, Arndt
  M. 2019  Quantum superposition of molecules beyond 25 {kDa}. {\em Nature
  Physics} \textbf{15}, 1242--1245.

\bibitem{Diosi21}
Donadi S, Piscicchia K, Curceanu C, Di{\'o}si L, Laubenstein M, Bassi A. 2021
  Underground test of gravity-related wave function collapse. {\em Nature
  Physics} \textbf{17}, 74--78.

\bibitem{Nucleon2001}
Thomas AW, Weise W. 2001 {\em The Structure of the Nucleon} vol.~1.
Wiley Online Library.

\bibitem{Bass2005}
Bass SD. 2005  The spin structure of the proton. {\em Reviews of Modern
  Physics} \textbf{77}, 1257--1302.

\bibitem{Bass2013}
Aidala CA, Bass SD, Hasch D, Mallot GK. 2013  The spin structure of the
  nucleon. {\em Reviews of Modern Physics} \textbf{85}, 655--691.

\bibitem{Spin21}
Ji X, Yuan F, Zhao Y. 2021  What we know and what we don't know about the
  proton spin after 30 years. {\em Nature Reviews Physics} \textbf{3}, 27--38.

\bibitem{Kaons2012}
Hiesmayr BC, Di~Domenico A, Curceanu C, Gabriel A, Huber M, Larsson J{\AA},
  Moskal P. 2012  Revealing {B}ell's nonlocality for unstable systems in high
  energy physics. {\em The European Physical Journal C} \textbf{72}, 1856.

\bibitem{BK_ent}
Banerjee S, Alok AK, MacKenzie R. 2016  Quantum correlations in {$B$} and {$K$}
  meson systems. {\em The European Physical Journal Plus} \textbf{131}, 129.

\bibitem{EPR_quarks}
Tu Z, Kharzeev DE, Ullrich T. 2020  {E}instein--{P}odolsky--{R}osen Paradox and
  Quantum Entanglement at Subnucleonic Scales. {\em Physical Review Letters}
  \textbf{124}, 062001.

\bibitem{Ent_weak}
Iskander G, Pan J, Tyler M, Weber C, Baker O. 2020  Quantum entanglement and
  thermal behavior in charged-current weak interactions. {\em Physics Letters
  B} \textbf{811}, 135948.

\bibitem{QI_top}
Afik Y, de~Nova J. 2021  Entanglement and quantum tomography with top quarks
  at the {LHC}. {\em The European Physical Journal Plus} \textbf{136}, 907.

\bibitem{PTomography1}
Chuang IL, Nielsen MA. 1997  Prescription for experimental determination of the
  dynamics of a quantum black box. {\em Journal of Modern Optics} \textbf{44},
  2455--2467.

\bibitem{PTomography2}
Poyatos JF, Cirac JI, Zoller P. 1997  Complete Characterization of a Quantum
  Process: The Two-Bit Quantum Gate. {\em Physical Review Letters} \textbf{78},
  390--393.

\bibitem{CHSH}
Clauser JF, Horne MA, Shimony A, Holt RA. 1969  Proposed Experiment to Test
  Local Hidden-Variable Theories. {\em Physical Review Letters} \textbf{23},
  880--884.

\bibitem{AspectBell2015}
Aspect A. 2015  Closing the door on {E}instein and {B}ohr's quantum debate.
  {\em Physics} \textbf{8}, 123.

\bibitem{Cirelson}
Cirel'son BS. 1980  Quantum generalizations of {B}ell's inequality. {\em
  Letters in Mathematical Physics} \textbf{4}, 93--100.


\bibitem{Tomo2004}
Paris M, Rehacek J, editors. 2004 {\em Quantum State Estimation} vol. 649{\em
  Lecture Notes in Physics}.
Springer Science \& Business Media.

\bibitem{Zurek_decoh}
Zurek WH. 2003  Decoherence, einselection, and the quantum origins of the
  classical. {\em Review Modern Physics} \textbf{75}, 715--775.

\bibitem{Zyczkowski}
Bengtsson I, {\.Z}yczkowski K. 2020 {\em Geometry of Quantum States: an
  Introduction to Quantum Entanglement}.
Cambridge University Press 2 edition.

\bibitem{AATomo}
D'Ariano GM, Lo~Presti P. 2001  Quantum Tomography for Measuring Experimentally
  the Matrix Elements of an Arbitrary Quantum Operation. {\em Physical Review
  Letters} \textbf{86}, 4195--4198.

\bibitem{MarkovGPT}
Yoshida Y, Hayashi M. 2020  Asymptotic properties for Markovian dynamics in
  quantum theory and general probabilistic theories. {\em Journal of Physics A:
  Mathematical and Theoretical} \textbf{53}, 215303.

\bibitem{Helstrom_book}
Helstrom CW. 1976 {\em Quantum Detection and Estimation Theory}.
Academic Press.

\bibitem{Childs}
Childs AM, Young J. 2016  Optimal state discrimination and unstructured search
  in nonlinear quantum mechanics. {\em Physical Review A} \textbf{93}, 022314.

\bibitem{Gisin1989}
Gisin N. 1989  Stochastic quantum dynamics and relativity. {\em Helvetica
  Physica Acta} \textbf{62}, 363--371.

\bibitem{Grudka2015}
Grudka A, Horodecki M, Horodecki R, W\'ojcik A. 2015  Nonsignaling quantum
  random access-code boxes. {\em Physical Review A} \textbf{92}, 052312.

\bibitem{Fidelity}
Jozsa R. 1994  Fidelity for Mixed Quantum States. {\em Journal of Modern
  Optics} \textbf{41}, 2315--2323.

\bibitem{EHS2020}
Sakharwade N, Studzi\'nski M, Eckstein M, Horodecki P. 2021  On two instances
  of random access code in the quantum regime. \textit{In preparation}.

\bibitem{Postquantum_steering}
Sainz AB, Brunner N, Cavalcanti D, Skrzypczyk P, V\'ertesi T. 2015  Postquantum
  Steering. {\em Physical Review Letters} \textbf{115}, 190403.

\bibitem{Postquantum_steering2}
Sainz AB, Hoban MJ, Skrzypczyk P, Aolita L. 2020  Bipartite Postquantum
  Steering in Generalized Scenarios. {\em Physical Review Letters}
  \textbf{125}, 050404.

\bibitem{QNS}
Piani M, Horodecki M, Horodecki P, Horodecki R. 2006  Properties of quantum
  nonsignaling boxes. {\em Physical Review A} \textbf{74}, 012305.

\bibitem{Random_channels}
Kukulski R, Nechita I, Pawela {\L}, Pucha{\l}a Z, {\.Z}yczkowski K. 2021
  Generating random quantum channels. {\em Journal of Mathematical Physics} \textbf{62}, 062201.

\bibitem{Marletto20}
Marletto C, Vedral V. 2020  Witnessing nonclassicality beyond quantum theory.
  {\em Physical Review D} \textbf{102}, 086012.

\bibitem{Rijavec20}
Rijavec S, Carlesso M, Bassi A, Vedral V, Marletto C. 2021  Decoherence effects
  in non-classicality tests of gravity. {\em New J. Phys.} \textbf{23}, 043040.

\bibitem{NMR20}
Chekhovich EA, da~Silva SFC, Rastelli A. 2020  Nuclear spin quantum register in
  an optically active semiconductor quantum dot. {\em Nature Nanotechnology}
  \textbf{15}, 999--1004.

\bibitem{CollectiveNucl21}
Bocklage L, Gollwitzer J, Strohm C, Adolff CF, Schlage K, Sergeev I, Leupold O,
  Wille HC, Meier G, R{\"o}hlsberger R. 2021  Coherent control of collective
  nuclear quantum states via transient magnons. {\em Science Advances}
  \textbf{7}, eabc3991.

\bibitem{Interface_EN}
Gangloff D, {\'E}thier-Majcher G, Lang C, Denning E, Bodey J, Jackson D, Clarke
  E, Hugues M, Le~Gall C, Atat{\"u}re M. 2019  Quantum interface of an electron
  and a nuclear ensemble. {\em Science} \textbf{364}, 62--66.

\bibitem{ph_prep}
Mosley PJ, Lundeen JS, Smith BJ, Wasylczyk P, U'Ren AB, Silberhorn C, Walmsley
  IA. 2008  Heralded Generation of Ultrafast Single Photons in Pure Quantum
  States. {\em Physical Review Letters} \textbf{100}, 133601.

\bibitem{ph_prep_rev}
Meyer-Scott E, Silberhorn C, Migdall A. 2020  Single-photon sources:
  Approaching the ideal through multiplexing. {\em Review of Scientific
  Instruments} \textbf{91}, 041101.

\bibitem{e_prep}
Feist A, Echternkamp KE, Schauss J, Yalunin SV, Sch{\"a}fer S, Ropers C. 2015
  Quantum coherent optical phase modulation in an ultrafast transmission
  electron microscope. {\em Nature} \textbf{521}, 200--203.

\bibitem{SQUIRRELS}
Priebe KE, Rathje C, Yalunin SV, Hohage T, Feist A, Sch{\"a}fer S, Ropers C.
  2017  Attosecond electron pulse trains and quantum state reconstruction in
  ultrafast transmission electron microscopy. {\em Nature Photonics}
  \textbf{11}, 793--797.

\bibitem{PhaseShaping}
Specht HP, Bochmann J, M{\"u}cke M, Weber B, Figueroa E, Moehring DL, Rempe G.
  2009  Phase shaping of single-photon wave packets. {\em Nature Photonics}
  \textbf{3}, 469--472.

\bibitem{PacketShaping}
Ahmadian A, Malekfar R. 2021  Wave packet shaping for a single-photon source.
  {\em Journal of the Optical Society of America B} \textbf{38}, 783--791.

\bibitem{Gamma18}
Moskal P, Krawczyk N, Hiesmayr B, Ba{\l}a M, Curceanu C, Czerwi{\'n}ski E,
  Dulski K, Gajos A, Gorgol M, Del~Grande R et~al.. 2018  Feasibility studies
  of the polarization of photons beyond the optical wavelength regime with the
  {J-PET} detector. {\em The European Physical Journal C} \textbf{78}, 1--9.

\bibitem{Gamma19}
Hiesmayr BC, Moskal P. 2019  Witnessing entanglement in compton scattering
  processes via mutually unbiased bases. {\em Scientific Reports} \textbf{9},
  1--14.

\bibitem{Espin99}
Garraway BM, Stenholm S. 1999  Observing the spin of a free electron. {\em
  Physical Review~A} \textbf{60}, 63--79.

\bibitem{QSternGerlach}
Gallup GA, Batelaan H, Gay TJ. 2001  Quantum-Mechanical Analysis of a
  Longitudinal {S}tern--{G}erlach Effect. {\em Physical Review Letters}
  \textbf{86}, 4508--4511.

\bibitem{Espin20}
Mart\'{\i}nez-Cifuentes J, Fonseca-Romero KM. 2021  Spin-state estimation using
  the {S}tern--{G}erlach experiment. {\em Phys. Rev. A} \textbf{103}, 042202.

\bibitem{Trap2017}
Schneider G, Mooser A, Bohman M, Sch{\"o}n N, Harrington J, Higuchi T, Nagahama
  H, Sellner S, Smorra C, Blaum K et~al.. 2017  Double-trap measurement of the
  proton magnetic moment at 0.3 parts per billion precision. {\em Science}
  \textbf{358}, 1081--1084.

\bibitem{Tweezer14}
Kaufman A, Lester B, Reynolds C, Wall M, Foss-Feig M, Hazzard K, Rey A, Regal
  C. 2014  Two-particle quantum interference in tunnel-coupled optical
  tweezers. {\em Science} \textbf{345}, 306--309.

\bibitem{Beams2005}
Mane SR, Shatunov YM, Yokoya K. 2005  Spin-polarized charged particle beams in
  high-energy accelerators. {\em Reports on Progress in Physics} \textbf{68},
  1997--2265.

\bibitem{E_beams21}
Nie Z, Li F, Morales F, Patchkovskii S, Smirnova O, An W, Nambu N, Matteo D,
  Marsh KA, Tsung F, Mori WB, Joshi C. 2021  \textit{In Situ} Generation of
  High-Energy Spin-Polarized Electrons in a Beam-Driven Plasma Wakefield
  Accelerator. {\em Physical Review Letters} \textbf{126}, 054801.

\bibitem{pos_pol1}
Li YF, Chen YY, Wang WM, Hu HS. 2020  Production of Highly Polarized Positron
  Beams via Helicity Transfer from Polarized Electrons in a Strong Laser Field.
  {\em Physical Review Letters} \textbf{125}, 044802.

\bibitem{Gamma_pol1}
Tang S, King B, Hu H. 2020  Highly polarised gamma photons from electron-laser
  collisions. {\em Physics Letters B} \textbf{809}, 135701.

\bibitem{Gamma_pol2}
Li YF, Shaisultanov R, Chen YY, Wan F, Hatsagortsyan KZ, Keitel CH, Li JX. 2020
   Polarized Ultrashort Brilliant Multi-{GeV} $\ensuremath{\gamma}$ Rays via
  Single-Shot Laser-Electron Interaction. {\em Physical Review Letters}
  \textbf{124}, 014801.

\bibitem{MAMI10}
Bernauer JC. 2010 {\em Measurement of the elastic electron--proton cross
  section and separation of the electric and magnetic form factor in the $Q^2$
  range from 0.004 to 1 $(\mathrm{GeV}/c)^2$}.
PhD thesis Johannes Gutenberg-Universit\"at Mainz.

\bibitem{ep_JefLab04}
Christy ME, Ahmidouch A, Armstrong CS, Arrington J, Asaturyan R, Avery S, Baker
  OK, Beck DH, Blok HP, Bochna CW, Boeglin W, Bosted P, Bouwhuis M, Breuer H,
  Brown DS, Bruell A, Carlini RD, Chant NS, Cochran A, Cole L, Danagoulian S,
  Day DB, Dunne J, Dutta D, Ent R, Fenker HC, Fox B, Gan L, Gao H, Garrow K,
  Gaskell D, Gasparian A, Geesaman DF, Gu\`eye PLJ, Harvey M, Holt RJ, Jiang X,
  Keppel CE, Kinney E, Liang Y, Lorenzon W, Lung A, Markowitz P, Martin JW,
  McIlhany K, McKee D, Meekins D, Miller MA, Milner RG, Mitchell JH, Mkrtchyan
  H, Mueller BA, Nathan A, Niculescu G, Niculescu I, O'Neill TG, Papavassiliou
  V, Pate SF, Piercey RB, Potterveld D, Ransome RD, Reinhold J, Rollinde E,
  Roos P, Sarty AJ, Sawafta R, Schulte EC, Segbefia E, Smith C, Stepanyan S,
  Strauch S, Tadevosyan V, Tang L, Tieulent R, Uzzle A, Vulcan WF, Wood SA,
  Xiong F, Yuan L, Zeier M, Zihlmann B, Ziskin V. 2004  Measurements of
  electron-proton elastic cross sections for $0.4 < Q^{2} < 5.5
  (\mathrm{GeV}/c)^{2}$. {\em Phys. Rev. C} \textbf{70}, 015206.

\bibitem{ep_JefLab19}
Xiong W, Gasparian A, Gao H, Dutta D, Khandaker M, Liyanage N, Pasyuk E, Peng
  C, Bai X, Ye L et~al.. 2019  A small proton charge radius from an
  electron--proton scattering experiment. {\em Nature} \textbf{575}, 147--150.

\bibitem{ep_JefLab20}
Androi\'{c} D, Armstrong DS, Asaturyan A, Bartlett K, Beaufait J, Beminiwattha
  RS, Benesch J, Benmokhtar F, Birchall J, Carlini RD, Cornejo JC, Dusa SC,
  Dalton MM, Davis CA, Deconinck W, Dowd JF, Dunne JA, Dutta D, Duvall WS,
  Elaasar M, Falk WR, Finn JM, Forest T, Gal C, Gaskell D, Gericke MTW, Grames
  J, Gray VM, Grimm K, Guo F, Hoskins JR, Jones D, Jones MK, Jones RT,
  Kargiantoulakis M, King PM, Korkmaz E, Kowalski S, Leacock J, Leckey JP, Lee
  AR, Lee JH, Lee L, MacEwan S, Mack D, Magee JA, Mahurin R, Mammei J, Martin
  JW, McHugh MJ, Meekins D, Mei J, Mesick KE, Michaels R, Micherdzinska A,
  Mkrtchyan A, Mkrtchyan H, Morgan N, Narayan A, Ndukum LZ, Nelyubin V,
  Nuruzzaman, van Oers WTH, Owen VF, Page SA, Pan J, Paschke KD, Phillips SK,
  Pitt ML, Radloff RW, Rajotte JF, Ramsay WD, Roche J, Sawatzky B, Seva T,
  Shabestari MH, Silwal R, Simicevic N, Smith GR, Solvignon P, Spayde DT,
  Subedi A, Subedi R, Suleiman R, Tadevosyan V, Tobias WA, Tvaskis V,
  Waidyawansa B, Wang P, Wells SP, Wood SA, Yang S, Zang P, Zhamkochyan S. 2020
   Precision Measurement of the Beam-Normal Single-Spin Asymmetry in
  Forward-Angle Elastic Electron-Proton Scattering. {\em Phys. Rev. Lett.}
  \textbf{125}, 112502.

\bibitem{ep_control10}
Christy ME, Bosted PE. 2010  Empirical fit to precision inclusive
  electron-proton cross sections in the resonance region. {\em Phys. Rev. C}
  \textbf{81}, 055213.

\bibitem{ZeilingerAI}
Melnikov AA, Poulsen~Nautrup H, Krenn M, Dunjko V, Tiersch M, Zeilinger A,
  Briegel HJ. 2018  Active learning machine learns to create new quantum
  experiments. {\em Proceedings of the National Academy of Sciences}
  \textbf{115}, 1221--1226.

\bibitem{ZeilingerAI2}
Krenn M, Erhard M, Zeilinger A. 2020  Computer-inspired quantum experiments.
  {\em Nature Reviews Physics} \textbf{2}, 649--661.

\bibitem{NQM_NP}
Abrams DS, Lloyd S. 1998  Nonlinear Quantum Mechanics Implies Polynomial-Time
  Solution for $\mathit{NP}$-Complete and \#$\mathit{P}$ Problems. {\em
  Physical Review Letters} \textbf{81}, 3992--3995.

\bibitem{NTCC}
Brassard G, Buhrman H, Linden N, M\'ethot AA, Tapp A, Unger F. 2006  Limit on
  Nonlocality in Any World in Which Communication Complexity Is Not Trivial.
  {\em Physical Review Letters} \textbf{96}, 250401.

\bibitem{Aaronson2005}
Aaronson S. 2005  Guest column: {NP}-complete problems and physical reality.
  {\em ACM Sigact News} \textbf{36}, 30--52.

\bibitem{NANLC}
Linden N, Popescu S, Short AJ, Winter A. 2007  Quantum Nonlocality and Beyond:
  Limits from Nonlocal Computation. {\em Physical Review Letters} \textbf{99},
  180502.

\bibitem{InformationCausality}
Paw{\l}owski M, Paterek T, Kaszlikowski D, Scarani V, Winter A, {\.Z}ukowski M.
  2009  Information causality as a physical principle. {\em Nature}
  \textbf{461}, 1101--1104.

\bibitem{LorentzCPT_2011}
Kosteleck\'y VA, Russell N. 2011  Data tables for {L}orentz and {$CPT$}
  violation. {\em Reviews of Modern Physics} \textbf{83}, 11--31.

\bibitem{Lake2010}
Lake B, Tsvelik AM, Notbohm S, Tennant DA, Perring TG, Reehuis M, Sekar C,
  Krabbes G, B{\"u}chner B. 2010  Confinement of fractional quantum number
  particles in a condensed-matter system. {\em Nature Physics} \textbf{6},
  50--55.

\bibitem{Kormos2017}
Kormos M, Collura M, Tak{\'a}cs G, Calabrese P. 2017  Real-time confinement
  following a quantum quench to a non-integrable model. {\em Nature Physics}
  \textbf{13}, 246--249.

\end{thebibliography}
\end{document}